\documentclass[twocolumn,
preprintnumbers,amsmath,amssymb,prl,superscriptaddress]{revtex4}

\usepackage{amsmath,amssymb,amsthm}
\usepackage{graphicx}
\usepackage{dcolumn}

\newcommand {\beq}{\begin{eqnarray}}
\newcommand {\eeq}{\end{eqnarray}}

\newcommand{\scF}{\ensuremath{\mathcal{F}}}

\begin{document}

\preprint{CALT-68-2721,
IPMU09-0025,
UT-09-04}

\title{Emergent Calabi-Yau Geometry}

\author{Hirosi Ooguri}

\affiliation{California Institute of Technology, Pasadena, CA 91125, USA}
\affiliation{Institute for the Physics and Mathematics of the Universe,
University of Tokyo, Kashiwa, Chiba 277-8586, Japan}

\author{Masahito Yamazaki}

\affiliation{California Institute of Technology, Pasadena, CA 91125, USA}
\affiliation{Institute for the Physics and Mathematics of the Universe,
University of Tokyo, Kashiwa, Chiba 277-8586, Japan}
\affiliation{Department of Physics, University of Tokyo,
Hongo 7-3-1, Tokyo 113-0033,
Japan}


\begin{abstract}
We show how the smooth geometry of Calabi-Yau manifolds
emerges from the thermodynamic limit of the statistical mechanical model
of crystal melting defined in our previous paper \cite{OY}.
In particular, the thermodynamic partition function of molten crystals 
is shown to be equal to the classical limit of the partition function 
of the topological string theory by relating the Ronkin function of 
the characteristic polynomial of the crystal melting model to the 
holomorphic 3-form on the corresponding Calabi-Yau manifold. 
\end{abstract}

\pacs{11.25.Uv,....}
\maketitle


\noindent
{\bf Introduction}:~
Stringy effects and quantum gravity effects are expected to
modify our concept of space and time, and understanding such effects
quantitatively is an important step in the exploration of physics at the Planck scale. In the perturbative string theory,
where the string coupling constant $g_s$ is assumed to be 
small, the string length is longer than the Planck length. 
In this regime, stringy corrections to spacetime geometry 
kicks in first and quantum gravity effects are obscured. 
In the context of the topological string theory on ${\bf C}^3$, 
it was pointed out in Ref.\ \cite{ORV} that the partition function 
can be resummed, and the result can be
 expressed in terms of a statistical model of 
crystal melting for large $g_s$. In our previous paper 
\cite{OY}, we generalized this construction to an arbitrary toric 
Calabi-Yau 3-fold. In the crystal melting model, the Calabi-Yau geometry 
is discretized, and each atom of the crystal can be regarded as 
a fundamental unit of the geometry.
In this Letter, we will show how the smooth 
Calabi-Yau geometry emerges from the discrete structure of
the crystal melting model
in the thermodynamic limit, where $g_s \rightarrow 0$. 
The topological string theory is relevant for counting of microstates
of black holes in the superstring theory \cite{OSV}, and we expect
that our result sheds some light on quantum nature of spacetime in the
superstring theory also.

A toric Calabi-Yau 3-fold $M$ is a K\"ahler quotient of
${\bf C}^{F+3}$ by $U(1)^{\otimes F}$, and its mirror manifold $\widetilde{M}$
is defined by the polynomial equation \cite{Hori,HIV},
\beq
u v + P(z,w) =0, ~~~ (u,v) \in {\bf C},~~~ (z,w) \in {\bf C}^{\times}.
\label{mirror}
\eeq
Here $P(z,w)$ is a Newton polynomial of the form,
\beq
 P(z,w) = \sum_{i=1}^{F+3} c_i(t) z^{n_i} w^{m_i},
\eeq
and  $c_i(t)$'s are functions of the K\"ahler
moduli $t$ of the original toric 3-fold $M$. 
The exponents $(n_i, m_i) \in {\bf Z}^2$ correspond to
lattice points of the toric diagram. For example,
for the mirror of ${\cal O}(-1)+{\cal O}(-1)$ bundle over 
${\bf C}P^1$, $P(z,w)$ is given by
\beq
P(z,w) = 1 + z +w + e^{t} zw.
\eeq

In this Letter, we will show that the Newton polynomial $P(z,w)$
for the mirror of a Calabi-Yau manifold is identical to the characteristic 
polynomial of the corresponding dimer model, which is the partition 
function of the model on a torus. The relation between $P(z,w)$ and 
the characteristic polynomial had been discussed earlier in \cite{FHKV}. 
Here, we will prove their precise equality including the dependence on the 
moduli $t$.

The dimer model enters into our discussion because of its relevance to
counting of D brane bound states. 
Consider D0 and D2 branes with a single D6 brane on $M$. The 
low energy effective theory is a supersymmetric quiver quantum 
mechanics characterized by a quiver diagram on $T^2$ \cite{BT,FHKV}. 
In our previous paper \cite{OY}, we showed that the 
dimer model defined in Ref.\ \cite{Szendroi,MR} 
gives a generating function of the Witten indices for 
bound states in the quiver quantum mechanics \cite{footnote1}.
 
We also constructed a statistical model of crystal melting
equivalent to the dimer model. 
The crystal consists of atoms of different types, 
each of which corresponds to a node
of the quiver diagram for the quantum mechanics, 
and the chemical bonds between
the atoms are dictated by edges of the diagram. 
The quiver diagram is drawn on $T^2$. 
To construct the initial configuration of the crystal, 
we start with the universal covering of the diagram over the plane and
pile up atoms on nodes following the rules
prescribed in Ref.\ \cite{OY}. This initial configuration
 corresponds to a single D6 brane with no D0 and
D2 charges. Removing atoms from the crystal 
generates bound states with non-zero D0 and D2 charges. 
Such molten crystal configurations are in one-to-one
correspondence with perfect matchings of the dimer model
defined in Ref.\ \cite{MR}, 
as shown in the Appendix of Ref.\ \cite{OY}. 

It is reasonable to expect that a classical geometric picture emerges
in the limit of large D0 and D2 charges since it can represent
a large black hole in the superstring theory. The corresponding
thermodynamic limit in the dimer model was studied in Ref.\ \cite{KOS}. 
In this Letter, we will show that the partition function of the
dimer model evaluated in the thermodynamic limit is equal to 
the genus-$0$ limit of the partition function of the topological
string theory on $M$. The dimer model has been formulated to 
describe the non-commutative Donaldson-Thomas theory
\cite{Szendroi,MR}, while the topological string theory for
a general toric Calabi-Yau manifold is
equivalent to the commutative Donaldson-Thomas theory \cite{DT}, 
as shown by \cite{MOOP}.
A relation of our results to the 
wall-crossing formula \cite{KS,NN,Nagao,JMC}
between the non-commutative and commutative 
Donaldson-Thomas theories
is currently under investigation.

The emergence of Calabi-Yau geometry from the thermodynamic
limit has been observed in Ref.\ \cite{ORV} in the case of ${\bf C}^3$. 
In this Letter,
we make the connection sharper and more explicit 
by showing the direct connection between the partition functions 
of the crystal melting model and the topological theory
for a general toric Calabi-Yau 3-fold. 


\medskip
\noindent
{\bf Thermodynamic Limit of the Crystal Melting Model}:~
The main object of study in this Letter is the partition function,
\beq
Z = \sum_{q_0, q_A} \Omega(q_0, q_A) e^{- g_s q_0-t^A q_A},
\label{partitionfunction}
\eeq
where $\Omega(q_0, q_A)$ is the Witten index for
bound states of 
$q_0$ D0 branes and $q_A$ D2 branes on the $A$-th 2-cycle
($A = 1,...,\textrm{dim}\,H_2(M)$) with a single D6 brane on a toric Calabi-Yau
manifold. According to the dictionary in Ref.\ \cite{OY}, 
the Witten index is equal (up to a sign) to the number of molten
crystal configurations where $q_0$ is the total number of
atoms removed,
whereas the relative numbers of different types of atoms 
removed from the crystal are specified by $q_A$'s. 
Later we will identify $g_s$ as the topological string coupling constant
and $t^A$ as the K\"ahler moduli of the toric Calabi-Yau manifold $M$. 

The behavior of $Z$ for $g_s\rightarrow 0$ can be evaluated by using 
the result of Ref.\ \cite{KOS}. Consider a finite covering the original quiver 
diagram, $N$ times in one direction and $N$ times in another direction
on $T^2$. $N$ is introduced as an infrared regulator, and 
we will take $N \rightarrow \infty$ at the end of the computation
so that we have the dimer model on the plane ${\bf R}^2$.
The surface of the crystal 
is determined by the height function $h$ over the plane. To define
$h$, we start with the canonical perfect matching $m_0$ 
of the dimer model corresponding to the initial crystal configuration
with no D0 or D2 charges. For any other 
perfect matching $m$, the superposition of $m_0$ and $m$ gives
a set of closed loops on the dimer graph. If $m$ corresponds to
a bound state with finite D0 and D2 charges, $m$ and $m_0$ differ only
in a finite region on the graph. The function $h_i$ ($i$: node of the 
the $N\times N$ cover of the quiver) is defined so that it is $0$ 
far away from the region where $m$ and
$m_0$ differ, and it increases by $1$ every time we cross 
a closed loop as we move inside of the region. 
The corresponding molten crystal configuration is obtained
by removing $h_i$ atoms of the initial crystal over the node $i$. 
In particular,
\beq
 \sum_i h_i = q_0.
\eeq

To take the thermodynamic limit,  
it is useful to introduce the Cartesian coordinates $(x,y)$,
$0 \leq x, y \leq 1$, on the $N\times N$ covering of $T^2$.
In the limit where $g_s \ll 1$ and $1 \ll N$, 
the height $h_i$ becomes a smooth function
$h(x,y)$. We rescale the height function by the factor of 
$1/N$ so that 
\beq
  N^2 \int_0^1 dx dy ~h(x,y) = \frac{q_0}{N},
\label{rescaledh}
\eeq
to take into account the large $N$ scaling of the 
partition function discussed in Ref.\ \cite{Sheffield} and quoted as
 Theorem 2.1 in Ref.\ \cite{KOS}. 
The statistical weight in the thermodynamic limit
is given by an integral of a surface tension 
$\sigma(\partial h)$, which is a function of the gradient of $h$,
as \cite{footnote3}
\beq
  Z \sim \exp\left[ N^2\, {\rm max}_h\! \int_0^1 dx dy  \left[-
\sigma(\partial h)
- g_sN h(x,y)\right]\right]. 
\label{thermlimit}
\eeq
The integral of $g_s N h(x,y)$ in the exponent comes from 
the weight factor $e^{-g_s q_0}$ 
in (\ref{partitionfunction}), and we used (\ref{rescaledh}). 
In the thermodynamic limit, we 
look for a height function $h(x,y)$ which maximize the
exponent. 

To derive the macroscopic surface tension from the 
microscopic crystal melting model, we first define 
a characteristic polynomial $\widetilde{P}(z,w)$ as the sum of 
perfect matchings with weights assigned to edges of the dimer model 
on $T^2$ \cite{KOS}. We then define its Ronkin function $R(x,y)$  by
\beq
  R(x,y) = \int_0^{2\pi} \ln \widetilde{P}(e^{x+i\theta}, e^{y+i\phi}) 
\frac{d\theta
d\phi}{(2\pi)^2}.
\label{ronkin}
\eeq
 According to Theorem 3.6 of Ref.\ \cite{KOS}, the surface tension 
$\sigma(\partial_xh, \partial_y h)$ is the Legendre transform
of the Ronkin function with respect 
to $(s,t)=(\partial_x h, \partial_y h)$ as
\beq
R(x,y)=  {\rm max}_{s,t} \left[ -\sigma(s,t)
+xs + yt\right].
\label{legendre}
\eeq 

The first step in relating the dimer model to the topological
string theory is to show that
the characteristic polynomial $\widetilde{P}(z,w)$ of the dimer model 
is equal to the Newton polynomial $P(z,w)$ for the mirror Calabi-Yau 
manifold (\ref{mirror}),
\beq
   \widetilde{P}(z,w) = P(z,w).
\label{agree}
\eeq

According to Ref. \cite{FV}, 
there is a one-to-one correspondence between 
perfect matchings of the dimer model on $T^2$ and bi-fundamental fields 
of gauged linear sigma model appearing in the K\"ahler quotient 
construction of the toric Calabi-Yau manifold $M$. 
They are then related, by change of variables described in Ref.\ \cite{FV},
to lattice points of the toric diagram and to terms $z^{n_i}w^{m_i}$ in the
Newton polynomial (\ref{mirror}). This shows that
there is a one-to-one correspondence between terms in $P(z,w)$ and $\tilde{P}(z,w)$,
as pointed out by \cite{FHKV}.
Furthermore we can show that their coefficients agree.
The K\"ahler moduli of $M$ are
the Fayet-Iliopoulos (FI) parameters
of the quiver quantum mechanics. 
According to the dictionary 
between quiver gauge theories and dimer models \cite{BT},
FI parameters are associated with nodes of the quiver diagram, or equivalently
to faces of the dimer model.
Thus, we can identify the FI parameters with the magnetic fluxes through
faces of the dimer model, which parametrize the energy of each perfect 
matching of the dimer model \cite{KOS}. 
Each perfect matching appears in 
$\widetilde{P}(z,w)$ with the weight given by an exponential of the
fluxes. On the other hand, the Newton polynomial is a sum of
$z^{n_i}w^{m_i}$, each of which corresponds to a lattice point of the toric diagram
and is weighted by an exponential of the K\"ahler moduli $t$ \cite{Hori}.
We have verified that the weight for perfect matchings and
the weight for lattice points of the toric diagram agree, and this proves 
the identity (\ref{agree}). 

Combining (\ref{thermlimit}) and (\ref{legendre}) and discarding a term
in total derivative in $(x,y)$, which is justified by the subtraction
of the linear piece in $R(x,y)$ discussed in the next paragraph, 
we find
\beq
  Z \sim \exp\left[ N^2 \int dx dy R\left(
\frac{g_sN}{2} x, \frac{g_sN}{2} y\right) \right]. 
\eeq
By rescaling $(x,y)$ by the factor of $gN/2$, this becomes 
\beq
  Z \sim \exp\left[ {4 \over g_s^2} \int dx dy R\left(x, y\right) \right]. 
\label{largeN}
\eeq
Note that the $N$ dependence has disappeared except that the range of the $(x,y)$ integral has been rescaled by the factor of $gN/2$. For 
$N \rightarrow \infty$ with small but fixed $g_s$, we have an integral over
the whole $(x,y)$ plane. 

The integral (\ref{largeN}) in the large $N$ limit is divergent.
To identify and subtract the divergent part, 
it is convenient to introduce the concept of the
amoeba \cite{footnote2}
which is a subset of ${\bf R}^2$ defined by, 
\begin{equation}
\begin{split}
{\rm Amoeba} = \{ (x, y)\in {\bf R}^2 :~&P(e^{x +i\theta},
e^{y+i\phi})=0 \\
&{\rm ~for~some~}(\theta,\phi)\}.
\label{amoeba}
\end{split}
\end{equation}  
In the thermodynamic limit, the amoeba corresponds to the liquid
phase of the crystal \cite{KOS}. If there are no interior points in 
the toric diagram, the complement of the amoeba is the solid phase,
where the crystal retains its original shape. There, the Ronkin function
$R(x,y)$ is linear \cite{PR}. If there are interior lattice
points in the toric diagram,
the amoeba acquires holes, inside of which are in the gas phase,
where the Ronkin function is again linear but the slope of the crystal
surface is different from the original one. 
The integral (\ref{largeN}) becomes finite if
we subtract the linear piece of the Ronkin function
in the solid phase so that the partition function is normalized
to be $1$ for the initial crystal configuration. 


\medskip
\noindent
{\bf Topological String at Genus $0$}:~
Our next task is to compute the genus-$0$ topological string partition 
function $\scF_0$ of the toric Calabi-Yau manifold $M$ and compare it with the 
thermodynamic limit of the partition function of the crystal melting model (\ref{largeN}). 
For this purpose, it is convenient to use the mirror Calabi-Yau manifold
$\widetilde{M}$  
defined by the equation (\ref{mirror}) since
$\scF_0$ can be
evaluated by the classical period integral as,
\beq
\scF_0=\int_{\beta_0} \Omega,
\label{period}
\eeq
where $\Omega$ is the holomorphic 3-form on the mirror, 
and $\beta_0$ is the
Lagrangian 3-cycle which is the mirror of the 6-cycle filling 
the entire toric Calabi-Yau manifold.

According to the microscopic derivation of the mirror symmetry by 
Hori and Vafa \cite{Hori}, the sigma-model on the toric Calabi-Yau manifold
is equivalent to the Landau-Ginzburg model with the superpotential
\beq
 W(u, x, y) = e^u P(e^x,e^y).
\eeq
It was shown in Ref.\ \cite{HIV} that an integral of $e^W$ in a 3-dimensional
subspace of the $(u,x,y)$ plane can 
be transformed into a period integral of the holomorphic 3-form $\Omega$
on the mirror Calabi-Yau manifold. Thus, we should be able to evaluate
(\ref{period}) as an integral of $e^W$. To do so, we need to identify the
contour of the integral. 

Since $g_s$ and $t^A$'s in (\ref{partitionfunction}) are 
taken to be real in the
dimer model, the Newton polynomial $P(x,y)$ in our case is with real 
coefficients. The mirror manifold (\ref{mirror}) has the 
complex conjugation involution, and thus the fixed point set is a natural
candidate for $\beta_0$. In fact, following the mirror symmetry 
transformation as described in Ref.\ \cite{Hori}, we find that the 6-cycle 
of the original toric Calabi-Yau manifold $M$ corresponds to the real section in the
$(u, x, y)$ space in the mirror $\widetilde{M}$. Thus, we find
\begin{equation}
\scF_0=\int_0^{\infty} du \int_{-\infty}^\infty
 dx dy\, e^{e^u P(e^x,e^y)} 
= - \int dx dy \ln P(e^x,e^y).
\label{log}
\end{equation}
The divergent part of the integral in
the $(x,y)$-plane can be removed by subtracting
a linear term in $\ln P(e^x, e^y)$ 
for $x,y \rightarrow \infty$ as we did for
(\ref{largeN}). The integral \eqref{log} is almost equal to 
the exponent of the partition function 
(\ref{largeN}) of the crystal melting model, except that 
we do not have the averaging over the phases $(\theta,\phi)$ to
define the Ronkin function as in (\ref{ronkin}). It turns out that
the integral over $(x,y)$ in (\ref{log}) removes the dependence 
on $(\theta,\phi)$, and thus the averaging process is not necessary.

To see this, let us define a generalization of $\scF_0$ for an
integral of $(x,y)$ with arbitrary phases as
\beq
\scF(\theta,\phi) 
= \int dx dy \ln P(e^{x+i\theta},e^{y+i\theta}).
\eeq
Taking derivatives of $\scF$ with
respect to $(\theta,\phi)$, we find the integrand becomes
a total derivative in $(x,y)$ as in 
\begin{equation}
\begin{split}
&\left(\alpha {\partial\over \partial \theta}+
 \beta {\partial\over \partial \phi}\right)\scF  \\
&=\int dx dy \,i\left(\alpha {\partial\over \partial x} 
+\beta {\partial\over \partial y} \right) \ln P(e^{x+i \theta},e^{y+i\phi}).
\end{split}
\end{equation}
If we choose $(\alpha,\beta)$ so that it is not in the directions
of the tentacles of the amoeba (\ref{amoeba}), the
boundary term is removed by the regularization and we find $(\alpha \partial_\theta + 
\beta \partial_\phi)\scF=0$. 
Since $\alpha, \beta$ is arbitrary except in the directions of tentacles of amoeba, $\scF$ is independent of $(\theta,\phi)$ and
agrees with its average. Namely, 
\beq
\scF_0= - \int dx dy \ln P(x,y) = -\int dx dy\, R(x,y).
\eeq

Thus, we found that the thermodynamic limit of the partition function 
 of the crystal melting model given by (\ref{largeN}) is equal to $\exp(-\frac{4}{g_s^2} 
\scF_0)$, which is the genus-$0$ partition function of the topological
string theory. This is what we wanted to show.


\medskip
\noindent{\bf Acknowledgment}:~
We would like to thank Alexei Borodin, 
Kentaro Hori, Kentaro Nagao and Andrei Okounkov 
for stimulating discussions. We thank Mina Aganagic for her comment on the earlier
version of this Letter.
This work is supported in part 
by DOE grant DE-FG03-92-ER40701 and by the 
World Premier International Research Center Initiative of MEXT of Japan.
H.~O. is also supported in part by a Grant-in-Aid for Scientific 
Research (C) 20540256 
of JSPS and by the Kavli Foundation. M.~Y. is also supported in part 
by JSPS 
and GCOE for Phys. Sci. Frontier of MEXT of Japan.



\begin{thebibliography}{99}

\bibitem{OY}
  H.~Ooguri and M.~Yamazaki,
  Comm. Math. Phys. (to be published), arXiv:0811.2801 [hep-th].


\bibitem{ORV}
  A.~Okounkov, N.~Reshetikhin and C.~Vafa,
  arXiv:hep-th/0309208.

\bibitem{OSV}
  H.~Ooguri, A.~Strominger and C.~Vafa,
  Phys.\ Rev.\  D {\bf 70}, 106007 (2004)
  [arXiv:hep-th/0405146].


\bibitem{Hori}
  K.~Hori and C.~Vafa,
  arXiv:hep-th/0002222.

\bibitem{HIV}
  K.~Hori, A.~Iqbal and C.~Vafa,
  arXiv:hep-th/0005247.



\bibitem{FHKV}
  B.~Feng, Y.~H.~He, K.~D.~Kennaway and C.~Vafa,
  Adv.\ Theor.\ Math.\ Phys.\  {\bf 12}, 3 (2008)
  [arXiv:hep-th/0511287].

\bibitem{BT}
  A.~Hanany and K.~D.~Kennaway,
  arXiv:hep-th/0503149;
%
  S.~Franco, A.~Hanany, K.~D.~Kennaway, D.~Vegh and B.~Wecht,
  JHEP {\bf 0601}, 096 (2006)
  [arXiv:hep-th/0504110];
%
  S.~Franco, A.~Hanany, D.~Martelli, J.~Sparks, D.~Vegh and B.~Wecht,
  JHEP {\bf 0601}, 128 (2006)
  [arXiv:hep-th/0505211];
%
  K.~Ueda and M.~Yamazaki,
  arXiv:math.AG/0703267.


\bibitem{Szendroi}
  B.~Szendr\"oi,
  Geom.\ Topol.\  {\bf 12}, 1171 (2008)
  [arXiv:0705.3419 [math.AG]].

\bibitem{MR}
  S.~Mozgovoy and M.~Reineke,
  arXiv:0809.0117 [math.AG].

\bibitem{footnote1}
  See, K.~Larjo,
  arXiv:0902.0614 [hep-th], 
  for subsequent developments.

\bibitem{KOS}
  R.~Kenyon, A.~Okounkov and S.~Sheffield,
  arXiv:math-ph/0311005.


\bibitem{DT}
 S.~K.~Donaldson and R.~P.~Thomas, 
 in {\it The Geometric Universe: Science, Geometry and The Work of Roger
 Penrose}, (Oxford University Press, Oxford, 1998);
%
  R.~P.~Thomas, 
J.\ Diff.\ Geom. {\bf 54}, 367 (2000)
  [arXiv:math.AG/9806111].


\bibitem{MOOP}
D.~Maulik, A.~Oblomkov, A.~Okounkov
and R.~Pandharipande, arXiv:math.AG/0809.3976

\bibitem{KS}
  M.~ Kontsevich and Y.~ Soibelman, 
  arXiv:0811.2435v1 [math.AG].


\bibitem{NN}
  K.~Nagao and H.~Nakajima,
  arXiv:0809.2992 [math.AG].

\bibitem{Nagao}
 K.~Nagao,
  arXiv:0809.2994 [math.AG].

\bibitem{JMC}
  D.~L.~Jafferis and G.~W.~Moore,
  arXiv:0810.4909 [hep-th];
%
  W.~ Y.~Chuang and D.~L.~Jafferis,
  arXiv:0810.5072 [hep-th].


\bibitem{Sheffield}
  S.~Sheffield, Ph.D. Thesis, Stanford University, 2003.


\bibitem{footnote3}
  Here we only show the $g_s$ dependence explicitly 
  and the dependence on the K\"ahler moduli $t^A$ is in $\sigma$.




\bibitem{FV}
  S.~Franco and D.~Vegh,
  JHEP {\bf 0611}, 054 (2006)
  [arXiv:hep-th/0601063].


\bibitem{footnote2}
  See Ref.\ \cite{Mikhalkin} for a survey of aspects of amoebae, and Refs.\ \cite{amoeba} for their uses in other aspects of gauge theories.

\bibitem{Mikhalkin}
  G.~Mikhalkin, 
  arXiv:math.AG/0108225



\bibitem{amoeba}
  T.~Maeda and T.~Nakatsu,
  Int.\ J.\ Mod.\ Phys.\  A {\bf 22}, 937 (2007)
  [arXiv:hep-th/0601233];
%
  T.~Fujimori, M.~Nitta, K.~Ohta, N.~Sakai and M.~Yamazaki,
  Phys.\ Rev.\  D {\bf 78}, 105004 (2008)
  [arXiv:0805.1194 [hep-th]].


\bibitem{PR}
  L.~ I.~Ronkin, in Complex Analysis in Modern Mathematics (Russian), 239–251, (FAZIS, Moscow, 2001);
%
  M. ~Passare and H.~Rullg{\aa}rd,
  Duke Math.\ J.\ {\bf 121}. 481 (2004).





\end{thebibliography}
\end{document}